\documentclass[12pt,preprint,aps,showpacs,color,floatfix]{revtex4}
\usepackage{epsfig,color}
\usepackage{rotating}
\newcommand{\Figref}[1]{Fig.\ \ref{#1}}
\newcommand{\figref}[1]{fig.\ \ref{#1}}

\def\be{\begin{equation}}
\def\ee{\end{equation}}
\def\bea{\begin{eqnarray}}
\def\eea{\end{eqnarray}}

\begin{document}

\title{How can we explore the onset of deconfinement by experiment?}

\author {J. Aichelin$^1$\footnote{invited speaker}, H. Petersen$^2$, S. Vogel$^2$, M. Bleicher$^2$ }
\address{$^1$ SUBATECH,  Universit\'e de Nantes, EMN, IN2P3/CNRS \\
  4, Rue Alfred Kastler, 44070 Nantes Cedex 03, France \\
$^2$, Institute for Theoretical Physics, Wolfgang Goethe University
of Frankfurt, Germany }


\begin{abstract}
There is little doubt that Quantumchromodynamics (QCD) is the theory
which describes strong interaction physics. Lattice gauge
simulations of QCD predict that in the $\mu,T$ plane there is a line
where a transition from confined hadronic matter to deconfined
quarks takes place. The transition is either a cross over (at low
$\mu$) or of first order (at high $\mu$). It is the goal of the
present and future heavy ion experiment at RHIC and FAIR to study
this phase transition at different locations in the $\mu,T$ plane
and to explore the properties of the deconfined phase. It is the
purpose of this contribution to discuss some of the observables
which are considered as useful for this purpose.
\end{abstract}
\maketitle
\section{Introduction}
The behavior of hadrons in an environment of finite temperature and
density and the phase transition towards a deconfined phase in which
quarks and gluons are the dominant degrees of freedom is a central
topic of theoretical nuclear physics since many years. Detailed
calculations have been revealed that hadrons react quite differently
if the are brought in a dense and/or hot environment. Vector mesons
change their width but not their pole mass when they are brought
into a dense environment \cite{rapp} whereas for $K^+$ mesons a
substantial change of the pole mass is predicted \cite{korpa} but
the width remains small. At low temperature but high density $K^-$
cannot be treated anymore as quasi particles having a quite
complicated spectral function\cite{luhab}. The different behavior of
the different hadrons comes from their different interactions with
their environment but many details of these interactions at finite
density and temperature are not well known

Statistical calculations yield a chemical freeze out energy density
of $1.1 GeV/fm^3$ for finite chemical potentials, well below the
energy density predicted by lattice gauge calculation for the
transition towards the deconfined phas where all hadrons become
unstable. This deconfined phase is not a weakly interacting plasma,
as one has thought for quite a time, but a liquid which can be
described by hydrodynamics much better than ever expected. When
applied to the scenario of an expanding quark gluon plasma these
hydrodynamical calculations describe quite well the experimental
observations if they start out from a strongly anisotropic initial
state, caused by the geometry of the reaction partners, which
expands while keeping local equilibrium.

From all these calculations we have a qualitative understanding of
strongly interacting matter but from a quantitative understanding we
are as far away as from an experimental verification of the
theoretical predictions. The many body theory of hadrons in matter
is complicated and many details are neither experimentally
accessible nor theoretically known. Therefore theoretical
predictions differ quantitatively. Due to the limited computer
capacity also lattice gauge calculations have not converged yet to
an exact temperature value at which the phase transition takes
place. Even if in the next years progress will be made in the
theoretical approaches the ultimate goal is to verify the
predictions experimentally and to convert theoretical predictions
into experimental facts.

In order to explore the properties of strongly interacting matter
complicated experiments have been performed and designed - at RHIC,
LHC and FAIR - in which in one single heavy ion reaction several
hundred particles are registered in the detectors. When registered,
however, all particles have to have their free mass and therefore
one can only learn something about the properties of strongly
interaction matter at high density/temperature if one understands
the time evolution of the system between the high density phase and
the detection.

Several ideas have been launched to asses matter properties at high
density/temperature:

a) To measure resonances. The decay products reflect the particle
properties at the point of disintegration which may be at finite
density. If the decay products interact strongly these particles are
sensitive to moderate densities only because the resonance cannot be
identified if one of the decay products interacts another time.

b) To measure dilepton pairs. Because leptons do practically not
interact with the expanding matter they may carry information on
particles which have been disintegrated in a dense environment. This
we discuss in section II.

c) To measure collective observables as discussed in section III.

d) To measure particles which can only be produced at the beginning
of the interaction when the density is quite high because later the
available energy is too low. This is the subject of chapter IV.

In this contribution I will critically review the significance of
some experimental observables for the exploration of the high
density zone at the future FAIR energies.

To study the sensitivity of the different probes on the properties
of high density zone we employ the UrQMD model which has been
successfully used to describe many of the stable and unstable
particles observed at AGS and RHIC energies \cite{weber}. Details of
this model may be found in \cite{urqmd}.
\section{dileptons}
\begin{figure}[hbt]
\epsfig{file=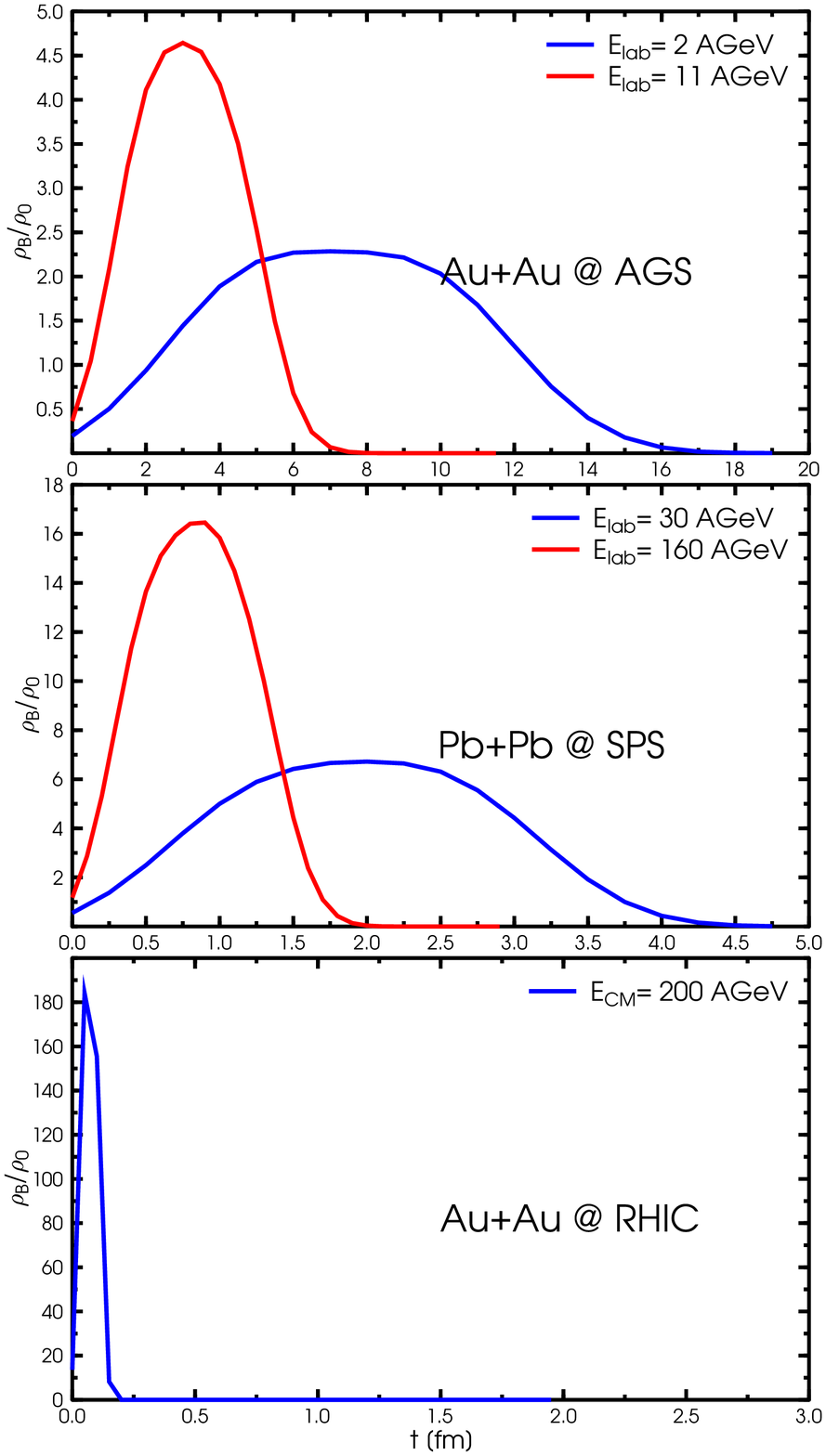,width=0.45\textwidth}
\epsfig{file=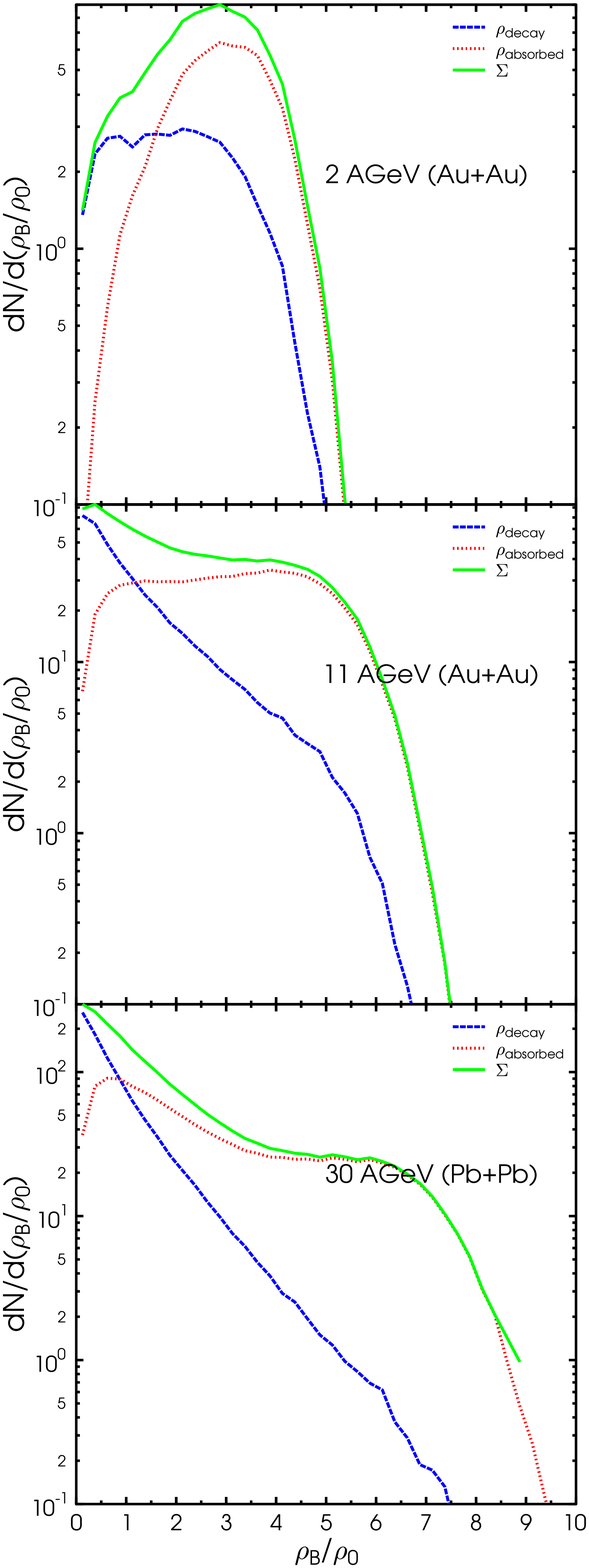,width=0.45\textwidth} \caption{Left: Time
evolution of the density of central heavy ion reactions for energies
ranging from $E_{lab}$=2 AGeV  $E_{cm}$=200 AGeV. Right:
Distribution of the density at which $\rho$ mesons disappear from
the system, either by reabsorption (dotted line) or by
disintegration (dashed line).} \label{bild1}
\end{figure}

Using the UrQMD model we studied the time evolution of the $\rho$
mesons which - due to their short life time - disintegrate while the
system is still in contact. Their decay products, especially the
dileptons, have been suggested as a possible source of information
on the high density zone of the reaction. In \Figref{bild1}, left,
we display the time evolution of the density as a function of time
for different energies, ranging from $E_{lab}$ = 2 AGeV (SIS) to
$E_{cm}$ = 200 AGeV (RHIC). We display the average density in the
rest system of the particles. Clearly, as expected, we see that with
increasing beam energy the maximal density of the system increases.
On the right hand side of the same figure we display the
distribution of the densities at the space-time points at which a
$\rho$ meson disappears during the reaction, either because it
decays (dashed line) or because it gets reabsorbed (dotted line). It
is evident that the higher the density the higher is the chance that
the $\rho$ meson becomes reabsorbed. Thus most of the $\rho$ mesons
which decay (and with a certain probability can be observed as a
dilepton pair in the detectors) are produced at a late time, long
after the system has passed the high density.
\begin{figure}[hbt]
\epsfig{file=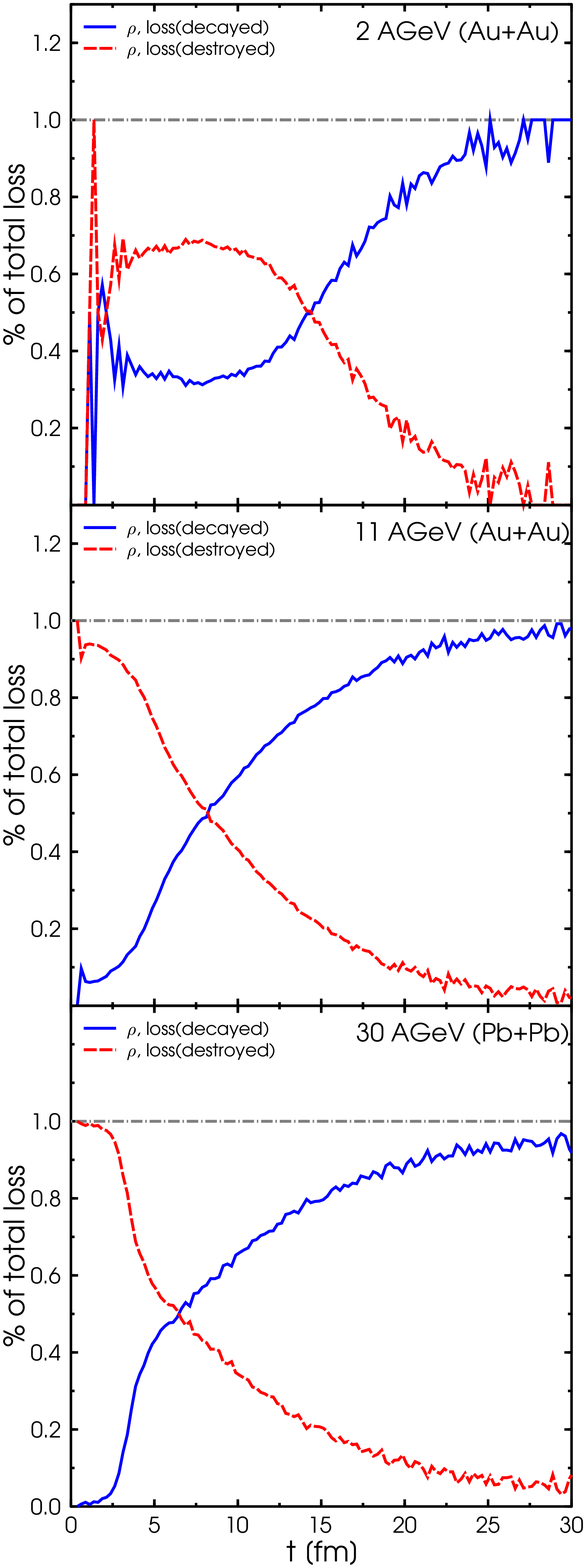,width=0.45\textwidth} \caption{
Fraction of the $\rho $ meson which decay and which get reabsorbed
(destroyed) as a function of time for 3 Beam energies between 2 AGeV
and 30 AGeV.} \label{bild2}
\end{figure}
It is clearly visible that the $\rho$ which disappear by decay come
from a very low densities, close or below normal nuclear matter
density. $\rho$ mesons which are produced at higher densities become
that fast reabsorbed that decay becomes a rare process. One can of
course discuss the details of this approach, especially the
properties of the $\rho$ at high density. The conclusion that
reabsorption and not decay is the dominant process at high densities
does not depend on these details. Therefore, dileptons coming from a
$\rho$ decay are not sensitive to system properties at high
densities. It is remarkable that the average density at the
disintegration point of the $\rho$ is at $E_{lab}$ = 30 AGeV even
lower than at $E_{lab}$ = 2 AGeV caused by the higher particle
multiplicity at higher energies. The fraction of $\rho$ mesons which
decay and of those which become reabsorbed we display in
\figref{bild2} as a function of time. Comparing \figref{bild1} and
\figref{bild2} we see that decay dominates only when the system is
dilute. Thus dileptons coming from resonance decays are sensitive to
system properties at low density only although they interact
exclusively by electromagnetic interactions.

\section{Collective Observables}
\begin{figure}[hbt]
\epsfig{file=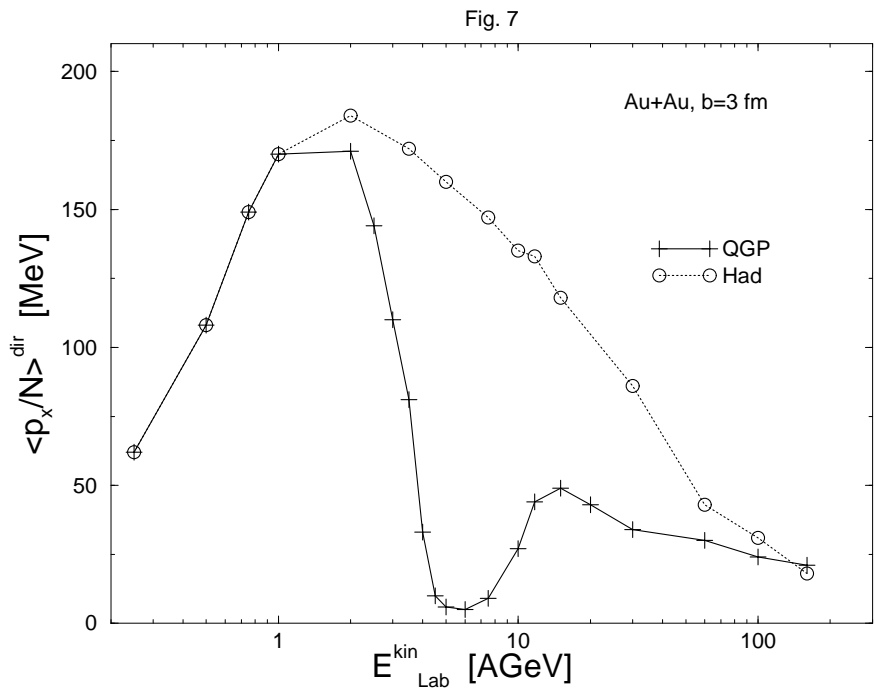,angle=270,width=0.7\textwidth}
\caption{The directed flow, $p_x^{\rm dir}$, as a function of beam
energy for Au+Au--collisions at $b=3$ fm. The full line (crosses)
corresponds to hydrodynamical calculations using the EoS with phase
transition, the dotted line (open circles) to those with the pure
hadronic EoS. From ref. \cite{risch}.} \label{bild3}
\end{figure}

\begin{figure}[hbt]
\epsfig{file=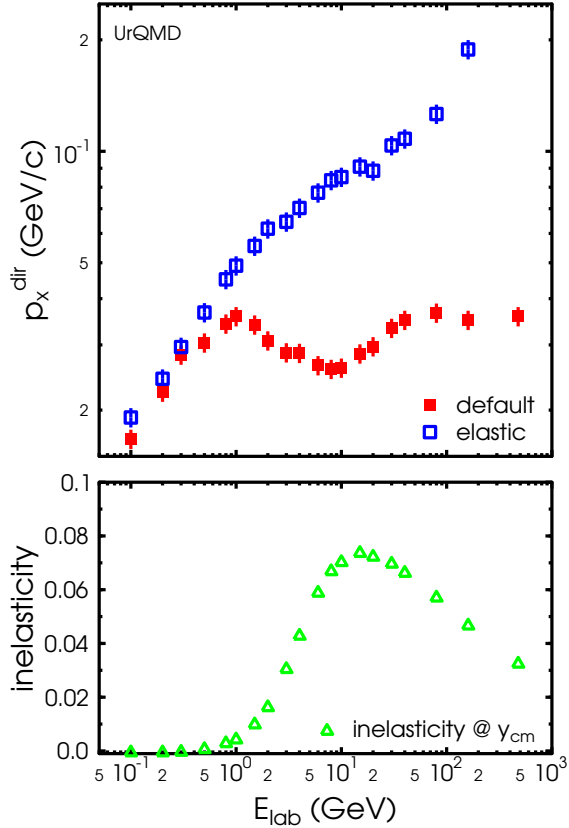,width=0.45\textwidth} \caption{Excitation
functions for central Au+Au (Pb+Pb) reactions. Top: Directed flow
$p_x^{\rm dir}$ of nucleons with only isotropic elastic interactions
(open squares) and with full elastic and inelastic collision term
(full squares). Bottom: Inelasticity (open triangles), from ref.
\cite{blei}} \label{bild4}
\end{figure}

As said, at the energies we are interested in the system is strongly
interacting. It is therefore possible that it acts collectively and
that collective observables carry information on the high density
state. Especially if the system passes the phase transition to
deconfined matter where (most of the) hadrons are not existing
anymore as stable particles collective observables are the only ones
which may carry a direct information. There are many collective
effects possible which are still explored. Here we concentrate on
one particular collective effect which has been identified in ref.
\cite{schu,risch} as a sign of the formation of a QGP. The phase
transition towards deconfined matter may soften the equation of
state. Such a softening would be visible in the excitation function
of the in-plane flow,
\begin{equation} p_x^{\rm dir} = \frac{1}{M}
\sum_i^M \, p_{x,i} \, {\rm sgn}(y_i),
\end{equation}
which decreases as a function of the beam energy much faster than
expected from an hadronic equation of state. For standard equations
of state this effect is maximal around the FAIR energies, where the
system is expected to reach the softest point, i.e. has the lowest
pressure to energy density ratio. \Figref{bild3} (from
ref.\cite{risch}) shows the excitation function  of $p_x^{\rm dir}$
in a hydrodynamical calculation. We see that after having reached a
maximum, $p_x^{\rm dir}$ decreases to a minimum if the system
becomes deconfined (QGP), whereas without the formation of a quark
gluon plasma (had) $p_x^{\rm dir}$ there is not such a minimum.
Thus measuring the excitation function of $p_x^{\rm dir}$ will bring
the presence of a quark gluon plasma to light. Unfortunately this
interpretation is laboring under a misapprehension. Using the more
elaborate UrQMD model in which local equilibrium is not enforced but
particles interact by known (free) cross sections we obtain the
excitation function of $p_x^{\rm dir}$ shown in \Figref{bild4}
\cite{blei}.
The reason for this form of the excitation function in UrQMD
calculations is the change of the angular distribution of the
nucleon-nucleon cross section with increasing energy and the
increasing probability that resonances are produced which decay
isotropically in their rest system. We see (top) that $p_x^{\rm
dir}$ increases with energy if the nucleon-nucleons cross section
were isotropic. The increasing anisotropy, seen in the NN data,
produces, however, a maximum of $p_x^{\rm dir}$ followed by a
decrease. At higher beam energies resonance production becomes
important which is measured by the inelasticity
\begin{equation}
{\rm Inelasticity}=\frac{\sum\, m_i}{E_{\rm total}}\quad {\rm at\,
y_{\rm cm}}\pm 0.5\quad.
\end{equation}
The isotropic decay of the resonances creates an increase of averge
transverse momentum of the particles in the system. The reabsorption
of the decay products depends on the azimuthal angle and causes an
observable increase of the in-plane flow $p_x^{\rm dir}$. These two
effects create in a realistic hadronic scenario an excitation
function of $p_x^{\rm dir}$ which resembles strongly that obtained
in hydrodynamical calculations if a quark gluons plasma is present.
The lesson to be learnt from these studies is that collective
observables in particular are complex and not easy to interpret and
that one has to be extremely carefully to identify an experimental
observation with one of the theoretically proposed reaction
scenarios before having excluded that others may lead to the same
predictions.
\section{Charmed Hadrons}
At SIS energies it has turned out that strange hadrons are a very
good tool to investigate the system at high density/temperature. The
reason for this is the fact that strange hadrons have to be produced
and that at SIS energies only in the initial phase, shortly after
projectile and target start to overlap, nucleon nucleons collisions
are sufficiently energetic to overcome the threshold
($\sqrt{s_{thres}}$= 2.548 GeV, corresponding to a beam energy of
1.583 GeV in pp collisions) for the production channel with the
lowest threshold ($NN\to K^+\Lambda N$). Once produced the $s$
quarks can still be exchanged between a baryon and a meson but the
probability that the $s$ and $\bar s$ quarks annihilate is
negligible. The charm multiplicity only gives information on the
high density zone because the threshold and hence the production
probability depends strongly on the properties of the strange
particles at the production point. The initial momentum distribution
is known from elementary collisions (and close to that expected from
three body phase space). One can therefore compare the initial and
final momentum distribution and use the difference to study the
interaction of the strange hadrons with the surrounding matter
during the expansion.

\begin{figure}
\epsfig{file=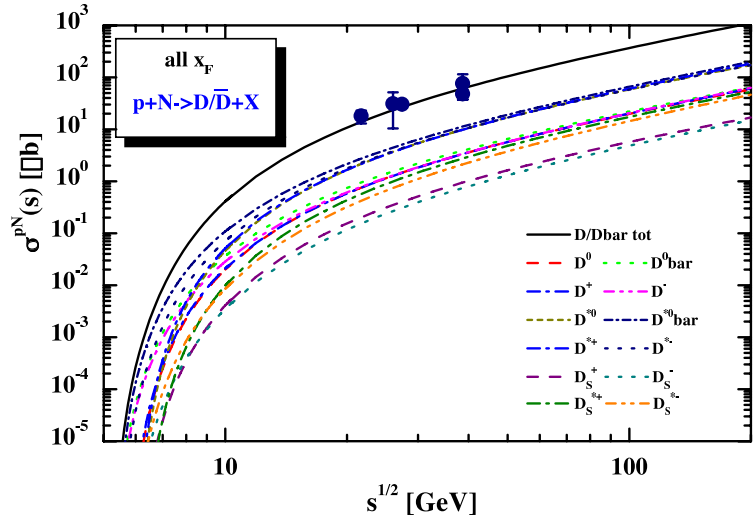,width=0.6\textwidth} \caption{ Cross section
parameterizations for open charm mesons in comparison to the
experimental data for $pp$. The upper solid lines denote the sum
over all $D/\bar{D}$ mesons. From ref.\cite{cas2}.} \label{bild6}
\end{figure}

It is certainly tempting and also planned to follow the same
strategy at FAIR energies by replacing strange hadrons by charmed
hadrons. At the highest FAIR energies ($E_{beam}$ = 30 AGeV,
corresponding to a center of mass energy of $\sqrt{s}= 7.74 \ GeV $
for a nucleon pair  we are slightly above threshold for charm
production process with the lowest threshold ($NN\to D^-(\bar
D^0)\Lambda_c N$, $\sqrt{s_{thres}}= 5.073(5.069) GeV)$ and
therefore - as the strange mesons at SIS energies - charmed hadrons
can only be produced initially in the high density zone. Before the
promising perspective to use charmed hadrons for a study of the high
density zone
\begin{figure}[!]
\epsfig{figure=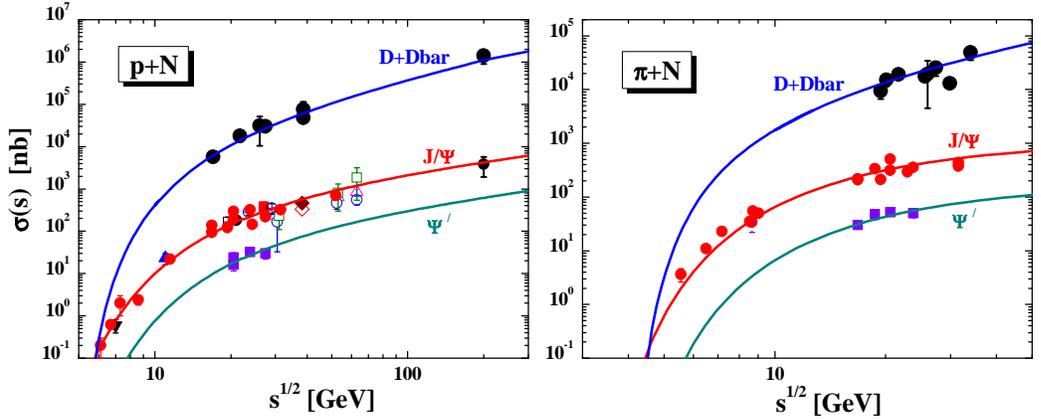,width=0.63\textwidth} \caption{The cross
section for $D+\bar D$, $J/\Psi$ and $\Psi^\prime$ meson production
in  $pN$ (left part) and $\pi N$ reactions (right part). The solid
lines show a parametrisations, whereas the symbols stand for the
experimental data. The $J/\Psi$ cross sections include the decay
from $\chi_c$ mesons. From ref.\cite{cas1}.} \label{xs_pp_pip}
\end{figure}
can lead to success a lot of work has to be accomplished. The
general problem is revealed in \Figref{xs_pp_pip} and \Figref{bild6}
which show the world data on charm production in elementary
collisions, compiled in ref. \cite{cas1,cas2}. On can see directly
that at the energies of interest at FAIR ($\sqrt{s} \approx 7 \
GeV)$ only $J/\psi$ production has been measured which is less
important at this energy  because this cannel has an higher
threshold than $NN\to D^-(\bar D^0)\Lambda_c N$. For the latter,
dominant, channel not a single data point is known. Well above
threshold  many channels contribute and the few existing data points
for $NN\to D^-(\bar D^0)+X$ are not of help to single out this cross
section. There is an additional problem, already known from $K^-$
physics at SIS. The $\Lambda_c$ will have a considerable charm
exchange cross section $\Lambda_c+ \pi \to D + N$ which is, however,
completely unknown. Due to this process the produced $c$ quarks will
be transferred to charmed mesons. Why is this of importance? All
charmed hadrons disintegrate before they reach the detector and
therefore one has to identify them by their decay products. The most
promising are energetic electrons and the $K^- \pi^+$ channel. The
branching ratio for disintegration into electrons of $\Lambda_c$
(4.5 \%) is much smaller than that of the corresponding $D^-$ meson
(17.2\%). Therefore, without knowing the repartition of the $c$
quark between mesons and baryons the observed electrons cannot be
used to determine the charm production multiplicity in a heavy ion
collision. This is also true, of course, for the $K^- \pi^+$ channel
which is only sensitive to the c-quark entrained in a meson.

This lack of knowledge on the production cross sections of charmed
hadrons in elementary collisions is also a very strong limitation
for any theoretical prediction for heavy ion collisions. Dynamical
simulation programs like UrQMD or HSD \cite{cas1,cas2} need these
cross sections as an input quantity. With the present knowledge of
these cross sections a reliable prediction for heavy ion collisions
at FAIR energies is impossible. Once these cross sections are known,
however, the excitation function of the multiplicity and hopefully
also the experimental momentum distribution of the charmed hadrons
which contain the desired information of the system properties at
high density and temperature can be analyzed and - there I am quite
sure - will reveal very interesting physics.

\end{document}